\documentclass[%
 reprint,
nofootinbib,
 amsmath,amssymb,
 aps,
]{revtex4-1}

\usepackage{graphicx}
\usepackage{dcolumn}
\usepackage{color}
\usepackage{bm}
\usepackage{caption,subcaption}
\usepackage{filecontents}

\begin{document}


\title{Exact results for the finite time thermodynamic uncertainty relation}

\author{Sreekanth K Manikandan}
\email{sreekanth.km@fysik.su.se}
\author{Supriya Krishnamurthy}%
 \email{supriya@fysik.su.se}
\affiliation{Department of Physics, Stockholm University,\\SE-10691 Stockholm, Sweden.}%

\date{\today}

\begin{abstract}
We obtain exact results for the recently discovered finite-time thermodynamic uncertainty relation, for the dissipated work $W_d$, in a stochastically driven system with non-Gaussian work statistics, both in the steady state and transient regimes, by obtaining exact expressions for any moment of $W_d$ at arbitrary times.
The uncertainty function (the Fano factor of $W_d$) is bounded from below by $2k_BT$ as expected, for all times $\tau$, in both steady state and transient regimes. The lower bound is reached at $\tau=0$ as well as when certain system parameters vanish (corresponding to an equilibrium state).  Surprisingly, we find that the uncertainty function also reaches a constant value at large $\tau$ for all the cases we have looked at. For a system starting and remaining in steady state, the uncertainty function   increases monotonically, as a function of $\tau$ as well as other system parameters, implying that the large $\tau$ value is also an upper bound. For the same system in the transient regime, however, we find that the uncertainty function can have a local minimum at an accessible time $\tau_m$, for a range of parameter values.
The large $\tau$ value for the uncertainty function is hence not a bound in this case. The non-monotonicity suggests, rather counter-intuitively, that there might be an optimal time for the working of microscopic machines, as well as an optimal configuration in the phase space of parameter values. Our solutions  show that the ratios of higher moments of the dissipated work are also bounded from below by $2k_BT$. For another model, also solvable by our methods, which never reaches a steady state, the uncertainty function, is in some cases, bounded from below by a value less than $2k_BT$.
\end{abstract}

\pacs{Valid PACS appear here}
\maketitle


Stochastic thermodynamics provides exact relations for thermodynamic quantities in mesoscopic or microscopic systems on experimentally accessible, finite time scales \cite{Seifert,expts}. One of the recent developments in the field is the discovery of the thermodynamic uncertainty relation \cite{biomol,hor1} which unifies two of the defining characteristics of non-equilibrium systems: non-zero dissipation and non-vanishing currents. For systems in a non-equilibrium steady state, the uncertainty relation provides a relation between the entropy production rate $\sigma$ in the steady state  and the ratio of the mean and variance of an arbitrary non vanishing current as,
\begin{align}
\label{gunc}
\frac{\text{Var}\left[X(\tau)\right]}{J^2\tau} &\geq \frac{2k_B}{\sigma}, & \left\langle X(\tau) \right\rangle =J\;\tau,
\end{align}
In this form, the uncertainty function captures a trade-off between precision (which is large when the variance is small) and the thermodynamic cost ($\sigma$). Specifying the integrated current to be equal to the dissipated work itself ($ \langle W_d \rangle =\sigma\tau T$, where $T$ is the temperature of the bath), we get the relation for the {\it Fano Factor} of the dissipated work,
\begin{equation}
\label{unceq}
\frac{\left\langle W_d^2\right\rangle-\left\langle W_d\right\rangle^2}{\left\langle W_d \right\rangle}\geq 2k_BT.
\end{equation}
We refer to the LHS of Eq. (\ref{unceq}) as the uncertainty function for conciseness.\par
The uncertainty relation was first conjectured in \cite{biomol} for bio-molecular processes, on the basis of an analysis performed within linear response theory for the steady state of multicyclic networks. Numerically though, it was seen to hold even beyond linear response \cite{biomol}. Since then,
a general proof has been provided \cite{hor1} by estimating a parabolic bound for the large-deviation rate function of steady-state currents for Markov jump processes. In \cite{univc}, different variations of such bounds
are discussed, for Markov processes with a finite number of states (the parabolic bound is however numerically shown to hold for any generic Markov process). A tighter bound for thermodynamically consistent currents in the steady state was obtained in \cite{tight}. Illustrations of these bounds at large-times have appeared in diverse contexts such as enzyme kinetics \cite{univb}, stochastic pumps \cite{pump}, first passage problems \cite{firstpass}, self propelled particles \cite{swim},  ballistic transport \cite{btr}, molecular motors \cite{motor}, biological motors \cite{cargo}, discrete time stochastic processes \cite{discr} and Brownian motion in a tilted potential \cite{tilt}. \par
The finite time version of the uncertainty relation, where the LHS of Eq. \ref{gunc} is studied as a function of $\tau$ at arbitrary times (but still in the steady state) is a very recent development. It was conjectured in \cite{unc} on the basis of extensive numerical evidence and a proof has only very recently been provided \cite{gin}, again in the framework of large deviation theory.
For a system not in steady-state, there is to our knowledge, only
the recent work of Pigolotti {\it et al} \cite{genL}, who, by introducing a random-time
transformation for Langevin processes, derive a stochastic differential equation for the entropy production rate, independent of the underlying model.
This results in an expression for the RHS of the uncertainty function
(in Eq. \ref{unceq}), in the steady state or a transient regime,
as a function of an {\it entropic time}. The large-time evaluation
of these expressions (using a Green-Kubo relation)
indicate that in the transient regime
the RHS of Eq. (\ref{unceq}) can sometimes be lower than $2k_BT$ \cite{genL}.
However solving for the entropic time as a function of the real time
is in general very complicated and the functional dependance of the uncertainty function on time is  only fully known for Gaussian work distributions \cite{Vanzone1,Vanzone2}.  Here the bound in Eq.\ \eqref{unceq} is exactly $2k_BT$ independent of $\tau$, and is a trivial consequence of the Fluctuation theorem (this connection is noted in \cite{tilt}). In this paper we compute, for the first time to our knowledge, exact results for the uncertainty function by calculating the arbitrary-time moment generating function (MGF) for a stochastically driven system having non-Gaussian work statistics. We obtain results using a path integral based approach \cite{skm},
for when the system is in the steady state as well as in a  transient regime. In both cases, the uncertainty function is seen to be bounded below by $2$ ( in units of $k_BT$). The bound is reached independently in the limits $\tau \rightarrow 0$ as well as when other dimensionless parameters $\rightarrow 0$, and can be understood in terms of the fluctuation theorem in the former case
and proximity-to-equilibrium in the latter case. At large times $\tau$, we find that the uncertainty function saturates to a value which depends on the system parameters in both cases. For arbitrary times, whether
the uncertainty function is monotonic or not depends
on initial conditions. For a system which starts and remains in steady state,
the function is strictly monotonic. However, for a system in the transient regime, the uncertainty function attains a local minimum at accessible times $\tau_m \ll \infty$, for a range of parameter values. This suggests very interestingly, that there might be an optimal time for microscopic machines to
function, as well an optimal configuration of parameter values. We also show using exact solutions for this system, that the ratio of higher moments of $W_d$ are also bounded from below by $2$, suggesting a hierarchy of bounds for steady state currents similar to the one appearing in Eq.\ \eqref{gunc}. We also study another model, the colloidal particle in the breathing parabola potential \cite{bretpjar,bretpprl,skm,Park}, which is a non-equilibrium system that never reaches a  steady state. We show, for a particular protocol that the uncertainty function attains a lower bound that is less than $2$. \par 
Our main results are for the \textit{sliding parabola} model which is a confined colloidal particle system, where the confining potential has the general form,
\begin{equation}
\label{slidp:pot }
V(x(t),\lambda(t))=\frac{1}{2}(x(t)-\lambda(t))^2.
\end{equation}
$x(t)$ is the position variable and $\lambda(t)$ is the externally modulated mean position. We have set the stiffness of the trap to $1$. In the overdamped limit, the dynamics of the colloidal particle in this potential can be described by the Langevin equation,
\begin{equation}
\label{slidp:x}
\dot{x}(t)=-\frac{1}{\tau_\gamma}\frac{\partial V(x,\lambda)}{\partial x(t)}+\sqrt{2D}\; \eta (t),
\end{equation}
where $\eta(t)$ is a thermal noise and $D$ is the diffusion coefficient. ${\tau }_{\gamma}$ is the relaxation time in the harmonic trap and is related to the temperature of the surrounding thermal environment $T$ by the Einstein relation $D\tau_\gamma=k_BT$. $\eta$ is assumed to be Gaussian with $\langle \eta(t) \rangle = 0$, and $\langle \eta(t)\; \eta(s) \rangle = \delta(t-s)$. In the case when $\lambda(t)$ is a deterministic driving protocol, the work distribution is known to be a Gaussian \cite{Vanzone1,Vanzone2,speck}. Eq.\ \eqref{slidp:x} has also been studied both experimentally \cite{expt} and analytically \cite{Sabhapandit,sabha1,Varley,skm}, when $\lambda(t)$ is a stochastic driving protocol\footnote{An equation of the form \eqref{slidp:x} can also be used to describe driven colloidal particles in active media. A recent experimental study on this can be found here \cite{nonb}.}. One of the cases studied is when $\lambda(t)$ is the Ornstein-Uhlenbeck process given by,
\begin{equation}
\label{slidp:OU}
\dot{\lambda}(t)=-\frac{\lambda(t)}{\tau_0}+\sqrt{2A}\; \xi (t).
\end{equation}
$\xi(t)$ is again assumed to be a Gaussian noise with $\langle \xi \rangle =0$ and $\langle \xi(t)\xi(s) \rangle=\delta(t-s)$. $\tau_0$ is the second natural time scale in the system in terms of the relaxation time of $\lambda$ correlations: $\langle \lambda(t) \lambda(s)\rangle=A\tau_0 \exp \left(-\frac{\vert t-s \vert}{\tau_0} \right)$. Notice that $A \tau_0$ could also be interpreted as an effective temperature \cite{temp}. The two noises are assumed to not have cross correlations, {\it i.e.}  $\langle \eta(t)\; \xi(s) \rangle =0$.  Equations \eqref{slidp:x} and \eqref{slidp:OU} together define the model we study here, which we refer to as the Stochastic Sliding parabola (SSP). The steady state probability distribution for the position variable $x(t)$ is known \cite{Sabhapandit}. One important parameter of the system, which measures how far the system is away from equilibrium \cite{expt,Sabhapandit} is given by, 
\begin{align}
\label{alpha}
\alpha=\frac{\left\langle x^2 \right \rangle_{st}}{\left\langle x^2 \right \rangle_{eq}} -1 = \frac{\theta \delta^2}{1+\delta}.
\end{align}
The subscripts {\it st} and {\it eq} stand for the steady-state and the equilibrium state\footnote{The equilibrium state corresponds to the situation when there is no external driving, {\it i.e.,} $\lambda(t)=0$. We have then, $\left\langle x^2 \right \rangle_{eq} = D\tau_\gamma$.} respectively. The dimensionless parameters $\theta$ and $\delta$ are defined as,
\begin{align}
\delta&=\frac{\tau_0}{\tau_\gamma},& \theta&=\frac{A}{D}.
\end{align}
Equilibrium is therefore defined as the limit $\delta \rightarrow 0$ and $\theta \rightarrow 0$. 
The large-time form of the MGF of $W_d$ for the SSP in the steady state has been studied previously \cite{Sabhapandit, Varley}. It is also known that the work distributions are non-Gaussian,  having exponentially decaying tails, with a power-law pre-factor \cite{skm}. In this paper, we are interested in arbitrary-time results for $W_d$ and hence  we begin first by demonstrating that the MGF of the dissipated work for the SSP,
is obtainable for arbitrary initial conditions and at arbitrary times as a function of all the parameters of interest ($\delta$, $\theta$ and $\tau_0$). \par

\begin{figure}[htb!]
 \centering
\includegraphics[scale=0.4]{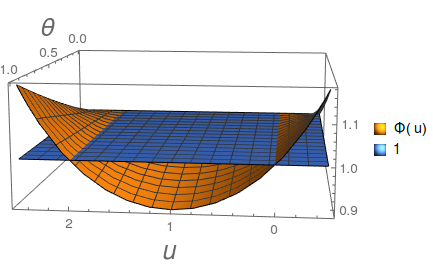}
 \caption{\label{figureone} $\Phi(u)$ as a function of $u$ and $\theta$ for $\delta =1 $ and $\tau_0 =1$ at $\tau=1$. $\Phi(0)=\Phi(2)=1$, which corresponds to probability-normalization and the integrated fluctuation theorem respectively.}
\end{figure}

We begin first by looking at a system, in a time interval $[0,\tau]$,
with a steady-state initial condition. The form of the dissipated work
$W_d$ is obtained from the ratio of the probabilities of forward and time-reversed trajectories \cite{Chernyk}. Using a path integral based approach \cite{pathintegral,skm}, we then write down the the MGF of $W_d$ upto a normalization constant $\textbf{C}$ as ( see supplementary material ),
\begin{align}
\begin{split}
\label{slidp:pwd}
\langle e^{-\frac{u}{2}\;W_d[x(\cdot),\;\lambda(\cdot)]}\rangle &=\textbf{C} \times \int dx_0 \;\int d\lambda_0\;\int dx_\tau\;\int d\lambda_\tau \\& \int_{x_0,\lambda_0}^{x_\tau,\lambda_\tau}\; \mathcal{D}[x(\cdot),\; \lambda(\cdot)]\;e^{-\; S[\;x(\cdot),\; \lambda(\cdot),\;u\;]},
\end{split}
\end{align}
where the action can be written down as,
\begin{align}
\begin{split}
S[\;x(\cdot),\; \lambda(\cdot),\;u\;]&=\frac{1}{4D}\;\left[\begin{array}{cc}x&\lambda
\end{array}\right]\;A_u\;\left[\begin{array}{c}x\\\lambda
\end{array}\right]\\ &+\text{ boundary terms}.
\end{split}
\end{align}
$A_u$ is a matrix differential operator \cite{functional,Kirsten}. Following the usual procedures for Gaussian integrations, the MGF at arbitrary times $\tau$, can be written down as a ratio of functional determinants,
\begin{equation}
\label{slidp:detr}
\langle e^{-\frac{u}{2}\; W_d\left[x(\cdot),\; \lambda(\cdot)\right]} \rangle_\tau = \sqrt{\frac{\det \textbf{A}_{u=0}}{\det \textbf{A}_{u}}}\equiv \Phi(u).
\end{equation}
This ratio of functional determinants may be obtained exactly using the methods elaborated in \cite{skm}. In Fig. \ref{figureone} we plot the exact solution of the generating function, as a function of $u$ and $\theta$. Notice that, $\Phi(u)$ is symmetric around $u=1$ as expected from the fluctuation theorem.
\par 
\begin{figure}[htb!]
  \centering
             \includegraphics[scale=0.4]{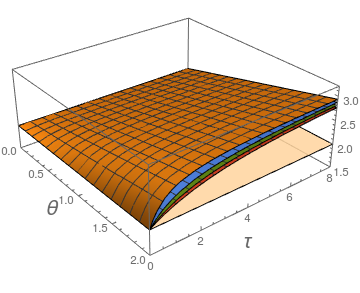}
            \caption{\label{newf}Unc($\tau$) in the steady state of the SSP for $\delta = 1$ and $\tau_0=1$ (Orange), $2$(Blue), $3$ (Green) and $4$ (Red). We have also plotted the plane Unc($\tau)$ $=2$.}
\end{figure}

Using standard techniques, various moments of the probability distribution may be extracted from Eq.\ \eqref{slidp:detr}. The first two moments are obtained as \footnote{The large deviation form of the moment generating function given in \citep{Varley} gives the same answer for the first moment. But for the second and higher order moments, only the leading order term in $\tau$ is obtained.},
\begin{align}
\label{exacts}
\left\langle W_d\right\rangle_\tau &= \frac{\delta ^2\; \theta  \;\tau }{(\delta +1) \;\tau_0},\\[5pt]
\label{exacts2}
\begin{split}
\left\langle W_d^2\right\rangle_\tau &= \scalebox{1}{$\frac{\delta ^2 \theta  \left(\delta ^2 (\delta +1)^2 \theta  \tau ^2+2 (\delta +1) \tau  \tau_0 \left(\delta ^2 (\theta +1)+2 \delta +1\right)-2 \delta ^2 \theta  \tau_0^2\right)}{(\delta +1)^4 \tau_0^2}$}\\
&+\scalebox{1}{$\frac{\delta ^4 \theta ^2 \left(\delta ^2 (\theta +1)-2 \delta +1\right) e^{-\frac{3 (\delta +1) \tau }{\tau_0}} \left(e^{\frac{(\delta +3) \tau }{\tau_0}}+e^{\frac{3 \delta  \tau +\tau }{\tau_0}}\right)}{\left(\delta ^2-1\right)^2 \left(\delta ^2 (\theta +1)+2 \delta +1\right)}$}\\
&-\scalebox{1}{$\frac{8 \delta ^7 \theta ^3 e^{-\frac{(\delta +1) \tau }{\tau_0}}}{(\delta -1)^2 (\delta +1)^4 \left(\delta ^2 (\theta +1)+2 \delta +1\right)}$}.
\end{split}
\end{align}
The uncertainty function ${\text{Unc}(\tau)}$ of dissipated work in the steady state (LHS of Eq. ~\eqref{unceq} where the averages are over the steady state distribution) for the SSP can now be computed analytically using Eq.\ \eqref{exacts} and \eqref{exacts2} (and is plotted in Fig. \ref{newf} for specific choices of the parameters). We find that for any value of $\tau$,
\begin{align}
2 \leq {\text{Unc}(\tau)} \leq 2\left(1+\alpha\right),
\label{eq:bounds}
\end{align}  
where $\alpha$ is exactly the expression in Eq.~(\ref{alpha}). The lower bound $2$ is obtained in the limit $\alpha \rightarrow 0$ (corresponding to the equilibrium limit ) as well as $\tau \rightarrow 0$ independently.
For non-zero values of $\alpha$, the function is monotonic in $\tau$
and reaches the value $2(1+ \alpha)$ for $\tau\rightarrow \infty$, leading to the inequalities in Eq.\ \eqref{eq:bounds}.
 \par
\begin{figure}[htb!]
  \centering
                \includegraphics[scale=0.4]{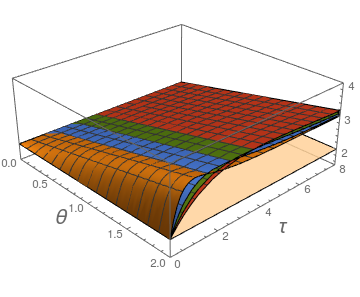}
           \caption{\label{figurefive}Unc($\tau$) for the SSP in the transient state, for $\delta=2$ and $\tau_0 =1$ (Orange), $2$ (Blue), $3$(Green) and $4$(Red). Unc($\tau$) is bounded from below by 2, but is not strictly monotonic.}
\end{figure}
\begin{figure}[htb!]
  \centering
        \includegraphics[scale=0.5]{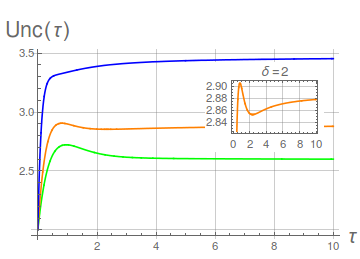}
            \caption{\label{figureseven} The uncertainty function has a local minimum for a range of values of $\delta$. The lines correspond to $\delta=1.2,\; 2, \text{ and } 6$ from bottom to top. The inset shows the minimum appearing at $\tau_m =2.34$ for $\delta =2$}
\end{figure}

Next, we study the SSP model in a transient regime, when starting from an equilibrium initial distribution, which we choose to be
\begin{align}
\label{tr:pini}
P(x_0,\lambda_0)=\sqrt{\frac{\delta}{\theta}}\frac{1}{2 \pi  D \tau_0}\exp\left(-\frac{\lambda_0^2}{2D\theta\tau_0}-\frac{\delta\left(x_0-\lambda_0\right)^2}{2D\tau_0}\right).
\end{align}
In this case, the dissipation function that satisfies a fluctuation theorem can be identified with the Jarzynski work \citep{skm}. We compute the exact moment generating function just as in the previous case (see supplementary material). It can again be verified that the uncertainty function is bounded from below by $2$ and the lower bound is attained in the limits $\alpha \rightarrow 0$ and $\tau \rightarrow 0$ independently. As $\tau \rightarrow \infty$, the uncertainty function attains the same constant
value as in Eq. \eqref{eq:bounds}. However for values of $\tau$ lying in between,
the function behaves non-monotonically and can take on values larger that the
value at $\tau \rightarrow \infty$.  In Fig.\ref{figurefive}, we plot
the uncertainty function in this regime, for specific choices of $\tau_0$ and $\delta$.
As a consequence of this non-monotonic behaviour, the uncertainty function displays a local minimum at an \textit{accessible} time $\tau_m \ll \infty$ for a range of parameter values. This is illustrated in Figure \ref{figureseven}. To understand further the dependence of $\tau_m$ on the other parameters, we have analysed the exact expression for the uncertainty function (provided in the supplementary material ). We have found that, for all range of parameters, $\theta$ and $\tau_0$, the minimum appears only between the critical values $1.367<\delta<3.890$. We have further verified this range of $\delta$ values using numerical simulations (which demonstrates an excellent agreement with the theory). We have also analysed the dependence of $\tau_m$ on the parameters $\theta$ and $\tau_0$. For fixed values of $\delta$ and $\tau_0$, $\tau_m$ is found to be independent of $\theta$. However, $\tau_m$ is seen to increase linearly with $\tau_0$ for fixed values of $\delta$ and $\theta$. Details of this analysis are provided in the supplementary material.\par  
\begin{figure}[htb!]
 \centering
 \includegraphics[scale=0.5]{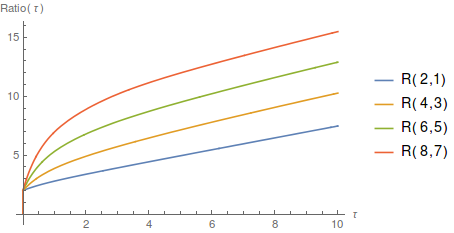}
 \caption{\label{hi} Ratio of $n$th moment of dissipated work to the $(n-1)$th ($\equiv R(n,n-1)$), for even values of $n$, in the steady state of the SSP. We have fixed $\delta=1$, $\theta=1$ and $\tau_0 =1$.}
 \end{figure}

 In the SSP model, both in the transient as well as in the steady state, we find that the higher order moments also satisfy the same lower bound in pairs. {\it i.e.,} the ratio of the 4th moment to the 3rd, the 6th to the 5th {\it etc}, are also bounded from below by 2, and the lower bound is attained as $\tau \rightarrow 0$, as well as in equilibrium. These ratios however do not seem to saturate for large $\tau$.
 We illustrate this in Figure. \ref{hi} for the steady state case. 
 The ratio of the cumulants are also always bounded from below. However, the limiting value as $\tau\rightarrow 0$ is not always 2.\par
\begin{figure}[htb!]
 \centering
 \includegraphics[scale=0.5]{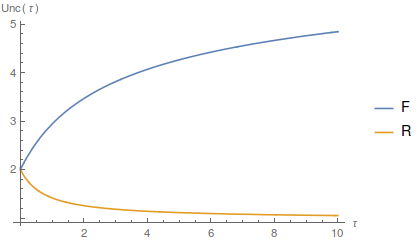}
 \caption{\label{brunc} Plot of uncertainty function w.r.t $\tau$, for the breathing parabola, forward(F) and reverse (R) protocols. For the reverse process, the uncertainty function is not bounded below by 2.}
 \end{figure}

 
For completeness of the discussion, we now study another model for which the system always remain in a transient state. The example we look at is the colloidal particle in a \textit{breathing parabola} potential, where the stiffness of the trap is changing in a time dependent manner according to a given protocol ( see supplementary material ). This system is also known to have non-Gaussian work statistics \cite{Nickelsen2011}. In figure \ref{brunc}, we plot the uncertainty function for both a specific forward protocol and the corresponding reverse protocol. We find again that the uncertainty function tends to $2$ as $\tau\rightarrow 0$. For the reverse process, the lower bound of the uncertainty function is $8/9$ and is less than $2$. For the forward process, the uncertainty function again saturates to an upper bound ($=8$) as $\tau\rightarrow \infty$.\par 
In summary, we have computed the exact form of the uncertainty function in a class of driven colloidal particle systems, having non-Gaussian work statistics, by computing the exact MGF of $W_d$ at arbitrary times $\tau$. Our results extend the list of solved problems for which the exact MGF is known and provides to our knowledge, the only non-trivial case where the uncertainty function can be solved exactly. We have looked at the SSP in
both the transient and steady-state regimes as well as the breathing
parabola which is always in a transient regime.
Our results show that in all cases, the uncertainty function $\rightarrow 2$ as $\tau \rightarrow 0$ ( also in the equilibrium limit for the SSP ) and saturates to a value for large $\tau$.
The large-$\tau$ value of the uncertainty function is however larger than $2$ only for systems that eventually reach a steady state. In this case it is also an upper bound in the case that the uncertainty function is monotonic in $\tau$.
In the transient regime of the SSP, the uncertainty function is non-monotonic and has a local minimum at an accessible time $\tau_m \ll \infty$, for a range of parameter values.
This is particularly interesting in connection with microscopic machines
in the biological context where the system presumably may be exactly in such a transient regime in many cases. Depending on what initial conditions the
system begins with, the position of the local minimum could provide an optimal
time-of-functioning beyond which the uncertainty function can only increase.
\section*{Acknowledgement}
We would like to thank Prof. Udo Seifert for commenting on the notes precursing this paper.

\begin{widetext}
\begin{center}
\section*{SUPPLEMENTARY MATERIAL}
\end{center}
This document provides further details of the calculations behind the results presented in the manuscript 'Exact results for the finite time thermodynamic uncertainty relation'. In Section \ref{slip:ss}, we present the derivation of  the exact MGF in the steady state of the SSP, and provide the exact form of the moments of $W_d$ in the transient case (the corresponding expressions in the steady state appear in the paper). In Section \ref{parm}, we provide the details of the calculations that determine the parameter dependence of $\tau_m$, the extremum of the uncertainty function that appears in the transient case. We present also a comparison with numerical simulations. In Section \ref{bpunc}, we present the exact results for the breathing parabola problem. \\
\section{ Exact calculation of the MGF of $W_d$ for the SSP model}
\label{slip:ss}
We look first at the steady state work statistics in the SSP model. Sampling the initial points from the stationary probability distribution for $x$ and $\lambda$ \cite{Sabhapandit}, 
\begin{align}
\label{slidp:pst}
p_{st}(x(t),\;\lambda(t))&=\frac{\exp \left(-\frac{(\delta+1) \left(\delta^2\theta (x-\lambda)^2+\delta \left(\theta x^2+\lambda^2\right)+\lambda^2\right)}{2D\tau_0\theta \left(\delta^2 (\theta+1)+2 \delta+1\right)}\right)}{2 \pi  \sqrt{\frac{D^2\tau_0^2\theta \left(\delta^2 (\theta+1)+2 \delta+1\right)}{\delta (\delta+1)^2}}}.
\end{align}
the MGF of $W_d$ can be written down in the following manner. First, the joint probability density functional of trajectories starting at $t=0$ at $(x_0,\lambda_0)$ and ending at $t=\tau$ at $(x_\tau,\lambda_\tau)$ may be written as,
\begin{equation}
\label{slidp:pxl}
P[x(\cdot),\lambda(\cdot)]=\textbf{N}\; \rm{exp}\; \bigg\lbrace-\int_{0}^{\tau}dt\;L(\dot{x}(t),x(t),\dot{\lambda}(t),\lambda(t),t)\bigg\rbrace
\end{equation}
with the Lagrangian,
\begin{equation}
\label{slidp:lagr}
L=\frac{1}{4D}\left(\;[\dot{x}+\frac{\delta(x-\lambda)}{\tau_0}]^2+\frac{1}{\theta}\;[\dot{\lambda}+\frac{\lambda}{\tau_0}]^2\right).
\end{equation}
The normalization constant for this case is \cite{pathintegral},
\begin{equation}
\label{slidp:norm}
\textbf{N}=\exp\left(\;\frac{1}{2}\;\left[\frac{\delta+1}{\tau_0}\right] \tau\;\right).
\end{equation}
The dissipated work in the steady-state in the time interval $[0,\tau]$ for the SSP is then,
\begin{equation}
\label{slidp:ep}
W_d\left[x,\lambda\right]=\frac{\delta}{D\tau_0}\int_0^\tau dt\; \lambda(t)\;\dot{x}(t)+\frac{\delta^2 \left(\delta \left(\theta \left(x_0^2-x_\tau^2\right)+2 x_0 \lambda_0-2 x_\tau \lambda_\tau-\lambda_0^2+\lambda_\tau^2\right)+2 x_0 \lambda_0-2 x_\tau \lambda_\tau\right)}{2 D \tau_0 \left(\delta^2 (\theta+1)+2 \delta+1\right)}.
\end{equation}
This form of the dissipated work can easily be obtained by equating it to the ratio of the probabilities of forward and time-reversed trajectories using Eq. (\ref{slidp:pst}) and Eq. (\ref{slidp:pxl}) and the form of the Lagrangian Eq. (\ref{slidp:lagr}). 
Hence, upto a normalization factor \textbf{C} (determined by Eq.\ \eqref{slidp:pst} and \eqref{slidp:norm}), we have the following expression for the MGF of $W_d$,
\begin{equation}
\label{slidp:pwd}
\langle e^{-\frac{u}{2}\;W_d[x(\cdot),\;\lambda(\cdot)]}\rangle=\textbf{C} \int dx_0 \;\int d\lambda_0\;\int dx_\tau\;\int d\lambda_\tau\int_{x(0),\lambda(0)=x_0,\lambda_0}^{x(\tau),\lambda(\tau)=x_\tau,\lambda_\tau}\; \mathcal{D}[x(\cdot),\; \lambda(\cdot)]\;e^{-\beta \; S[\;x(\cdot),\; \lambda(\cdot),\;u\;]},\;
\end{equation}
with the augmented action
\begin{equation}
\label{slidp:Sss}
\begin{split}
S[\;x(\cdot),\; \lambda(\cdot),\;u\;]&= \frac{(\delta+1) \left(\delta^2\theta (x_0-\lambda_0)^2+\delta \left(\theta x_0^2+\lambda_0^2\right)+\lambda_0^2\right)}{2D\tau_0\theta \left(\delta^2 (\theta+1)+2 \delta+1\right)}\\ &+ \int_{0}^{\tau}dt\;\frac{1}{4D}\left(\;[\dot{x}+\frac{\delta(x-\lambda)}{\tau_0}]^2+\frac{1}{\theta}\;[\dot{\lambda}+\frac{\lambda}{\tau_0}]^2\right)+\frac{u}{2}\;W_d[x,\lambda].
\end{split}
\end{equation}
after several partial integrations, it can be shown that the above quadratic action reduces to
\begin{equation}
\label{bcs}
S[\;x(\cdot),\; \lambda(\cdot),\;u\;]=\frac{1}{4D}\;\left[\begin{array}{cc}x&\lambda
\end{array}\right]\;\textbf{A}_u\;\left[\begin{array}{c}x\\\lambda
\end{array}\right]+\text{Boundary terms in } (x, \lambda, u),
\end{equation}
where the kernel is defined by the operator: 
\begin{align}
\label{slidp:ele}
\textbf{A}_u &= \left[\begin{array}{cc}-\frac{d^2}{d 
t^2}+\frac{\delta^2}{\tau_0^2} & k\;\frac{\delta}{\tau_0}\frac{d}{d t}-\frac{\delta^2}{\tau_0^2}\\-k\;\frac{\delta}{\tau_0}\frac{d}{d t}-\frac{\delta^2}{\tau_0^2} & -\frac{1}{\theta}\frac{d^2}{d 
t^2}+\frac{1}{\theta \tau_0^2}+\frac{\delta^2}{\tau_0^2} \end{array} \right];& k&\equiv 1-u.
\end{align} 
Carrying out the Gaussian integral, and requiring the boundary terms to vanish, the generating function at arbitrary times $\tau$ can be written down as a ratio of functional determinants,
\begin{equation}
\label{slidp:detr1}
\langle e^{-\frac{u}{2}\; W_d\left[x(\cdot),\; \lambda(\cdot)\right]} \rangle_\tau = \sqrt{\frac{\det \textbf{A}_{u=0}}{\det \textbf{A}_{u}}}\equiv \Phi(u).
\end{equation}
This ratio can be computed using a technique described in \cite{Kirsten} and recently used in \cite{skm}, which is based on the spectral -$\zeta$ functions of Sturm-Liouville type operators. Applying this method to the systems we study in this paper, it can be shown that this ratio can be obtained in terms of a characteristic polynomial function $F$ as,
\begin{align}
\label{mgf}
\left\langle e^{-\frac{u}{2}\; W_d\left[x(\cdot)\right]}\right\rangle_\tau &=\sqrt{\frac{F(1)}{F(k)}},  &F(k) &\equiv \text{Det}\left[M+N H(\tau)\right],
& k&=1-u,
\end{align}
where $H$ is the matrix of suitably normalized fundamental solutions of the homogeneous equation $\textbf{A}_u\; \vec{x}=0 $, and is defined as,
\begin{align}
H(t)&=
\left[
\begin{array}{cccc}
 x_1(t) & x_2(t) & x_3(t) & x_4(t) \\
 \lambda_1(t) & \lambda_2(t) & \lambda_3(t) & \lambda_4(t)\\
 \dot{x}_1(t) & \dot{x}_2(t) & \dot{x}_3(t) & \dot{x}_4(t)\\
\dot{\lambda}_1(t) & \dot{\lambda}_2(t) & \dot{\lambda}_3(t) & \dot{\lambda}_4(t) \\
\end{array}
\right],& H(0) &=\textbf{I}_4.
\end{align}
$M$ and $N$ have information about the boundary conditions from Eq.\ \eqref{bcs} and we require,
\begin{align}
M \left[\begin{array}{c}\vec{x}(0)\\\dot{\vec{x}}(0)\end{array}\right]&=0,&N \left[\begin{array}{c}\vec{x}(\tau)\\\dot{\vec{x}}(\tau)\end{array}\right]&=0.
\end{align}
A derivation of Eq.\ \eqref{mgf}, applicable to a class of driven Langevin systems with quadratic actions is given in \cite{skm}. We would also like to stress that, the expression given in Eq.\ \eqref{mgf} is valid only for  $u \in\left[u^-(\tau),u^+(\tau)\right] $ for which the operator $A_u$ doesn't have negative eigenvalues. The MGF is not analytic outside this interval.\par
For the SSP in the steady state, we find,
the four independent solutions of $A_u \vec{x}=0$ to be \begin{align}
\vec{x}_i&=\left[\begin{array}{c}x_i(t)\\\lambda_i(t)
\end{array}\right],&i&=1\text{ to }4,
\end{align} 
where,
\begin{align}
\lambda_i(t)&=\scalebox{1}{$\exp \left(\pm\;\frac{\tau  \sqrt{\frac{\delta ^2 \theta +\delta ^2+\delta ^2 \theta  \left(-(1-u)^2\right)\;\pm\;\tau_0^2 \sqrt{\frac{\delta ^4 \left(\theta -\theta  (1-u)^2+1\right)^2-2 \delta ^2 \left(\theta  \left((1-u)^2-1\right)+1\right)+1}{\tau_0^4}}+1}{\tau_0^2}}}{\sqrt{2}}\right)$},  \\
x_i(t)&=\scalebox{1}{$\frac{\tau_0 \left((u-1) \lambda_i '(t) \left(\delta ^2 \theta  (u-2) u-1\right)+\tau_0 \left(\delta  \lambda_i ''(t)+\tau_0 (u-1) \lambda_i ^{(3)}(t)\right)\right)+\delta  \lambda_i (t) \left(\delta ^2 \theta  (u-2) u-1\right)}{\delta ^3 \theta  (u-2) u}$}
\end{align}
Matrices $M$ and $N$ are given by,
\begin{align}
M&=\scalebox{1}{$\left(
\begin{array}{cccc}
 \frac{(1-(1-u) \theta ) \delta ^3+2 \delta ^2+\delta }{2 D \left((\theta +1) \delta ^2+2 \delta +1\right) \tau_0} & \frac{\delta  \left(((1-u) \theta -1) \delta ^2-u \delta -u+1\right)}{2 D \left((\theta +1) \delta ^2+2 \delta +1\right) \tau_0} & -\frac{1}{2 D} & 0 \\
 -\frac{(2-u) \delta ^2 (\delta +1)}{2 D \left((\theta +1) \delta ^2+2 \delta +1\right) \tau_0} & \frac{(2-u) \theta  \delta ^3+(\theta +1) \delta ^2+2 \delta +1}{2 D \theta  \left((\theta +1) \delta ^2+2 \delta +1\right) \tau_0} & 0 & -\frac{1}{2 D \theta } \\
 0 & 0 & 0 & 0 \\
 0 & 0 & 0 & 0 \\
\end{array}
\right)$}\\
N&=\scalebox{1}{$\left(
\begin{array}{cccc}
 0 & 0 & 0 & 0 \\
 0 & 0 & 0 & 0 \\
 \frac{\delta  \left(((1-u) \theta +1) \delta ^2+2 \delta +1\right)}{2 D \left((\theta +1) \delta ^2+2 \delta +1\right) \tau_0} & -\frac{\delta  \left((1-u) \theta  \delta ^2+\delta ^2+(1-u) \delta +\delta -u+1\right)}{2 D \left((\theta +1) \delta ^2+2 \delta +1\right) \tau_0} & \frac{1}{2 D} & 0 \\
 -\frac{u \delta ^2 (\delta +1)}{2 D \left((\theta +1) \delta ^2+2 \delta +1\right) \tau_0} & \frac{(\theta -(1-u) \theta ) \delta ^3+(\theta +1) \delta ^2+2 \delta +1}{2 D \theta  \left((\theta +1) \delta ^2+2 \delta +1\right) \tau_0} & 0 & \frac{1}{2 D \theta } \\
\end{array}
\right)$}
\end{align} \par
Using these, the MGF can be computed exactly using Eq.\ \eqref{mgf}, and various moments of the probability distribution can also be exactly obtained for any $\tau$. The exact expression of the first two moments for this case are given in the main text. Earlier calculations for the MGF obtained it only in the large-$\tau$ limit \citep{Varley}. The moments obtained from this large-$\tau$ form result in the same expression as we get for the first moment, but only
give the leading order term in $\tau$ for second and higher moments.\par
For the transient case of the SSP, the dissipation function which satisfies the fluctuation theorem can be identified with the Jarzynski work, 
\begin{equation}
\label{slidptr:ep}
W\left[x,\lambda\right]=\frac{\delta}{D\tau_0}\int_0^\tau dt\; \left(\lambda(t)-x(t)\right)\;\dot{\lambda}(t).
\end{equation}
The path integral calculation used in the steady-state case can be extended in a similar manner to this situation also. The first two moments of the transient work distribution of the SSP can be exactly computed and are given by the expressions,
\begin{align}
\label{exact}
\left\langle W\right\rangle_{\tau} &= \frac{\delta  \theta  \left(\frac{\delta ^2 \tau +\delta  \tau +\tau_0}{\tau_0}-e^{-\frac{(\delta +1) \tau }{\tau_0}}\right)}{(\delta +1)^2},\\
\begin{split}
\left\langle W^2\right\rangle_{\tau,\;\delta\geq 1} &=\frac{\delta  \theta }{(\delta +1)^4 \tau_0^2}\left(\delta ^5 \theta  \tau ^2+2 \delta ^2 \tau_0 ((\theta +3) \tau +2 \theta  \tau_0+\tau_0)+2 \delta  \tau_0 ((\theta +2) \tau_0+\tau )+2 \tau_0^2\right)\\
&+\frac{\delta  \theta }{(\delta +1)^4 \tau_0^2}\left(2 \delta ^4 \tau  (\theta  (\tau +\tau_0)+\tau_0)+\delta ^3 \left(\theta  \left(\tau ^2+4 \tau  \tau_0-2 \tau_0^2\right)+6 \tau  \tau_0\right)\right)\\
&-\frac{2 \delta  \theta  e^{-\frac{(\delta +1) \tau }{\tau_0}} \left(3 \delta ^5 \theta  \tau +\delta ^4 (-3 \theta  \tau +8 \theta  \tau_0+\tau_0)+3 \delta ^3 \theta  (\tau_0-\tau )+\delta ^2 (3 \theta  \tau +2 \theta  \tau_0-2 \tau_0)+3 \delta  \theta  \tau_0+\tau_0\right)}{(\delta -1)^2 (\delta +1)^4 \tau_0}\\
&\frac{\delta ^2 \theta ^2 e^{-\frac{6 (\delta +1) \tau }{\tau_0}} \left((\delta +1)^4 e^{\frac{2 (2 \delta +3) \tau }{\tau_0}}+(\delta +1)^4 e^{\frac{2 (3 \delta +2) \tau }{\tau_0}}+2 (\delta -1)^2 e^{\frac{4 (\delta +1) \tau }{\tau_0}}\right)}{(\delta -1)^2 (\delta +1)^4}
\end{split}
\end{align}
The expression for $\left\langle W^2 \right\rangle$ for $\delta <1$ is the same as above except for an overall minus sign. The above two expressions can be used to calculate the uncertainty function at
any value of $\tau$. It is interesting to note that though the above expressions seem to have
terms independent of $\tau$, both $\left\langle W_d \right\rangle$ and $\left\langle W_d^2 \right\rangle$ vanish at $\tau=0$ as they should. In addition, it is possible to show
that the limiting value of the uncertainty function is $2$ at
$\tau \rightarrow 0$, $\theta\rightarrow 0$ as well as $\delta \rightarrow 0$.
It is also easy to see that at $\tau \rightarrow \infty$, the uncertainty
function saturates to the same value $2 (1+\alpha )$ as in the steady-state case.

\subsection{Time of {\it minimum} uncertainty, $\tau_m$}
\label{parm}
For small times $\tau$, the uncertainty function is non-monotonic for the transient case. It increases at small $\tau$ upto a maximum value, followed by a minimum at $\tau_m$, beyond which it increases to the saturation value. This behaviour is however present only for a certain range of parameter values. In Figure \ref{deltar}, we plot the time derivative of the uncertainty function with respect to $\tau$, as a function of $\delta$ and $\tau$. The second point along the $\tau$ axis at which this function vanishes, corresponds to $\tau_m$.\par
\begin{figure}[htb!]
 \centering
 \includegraphics[scale=0.5]{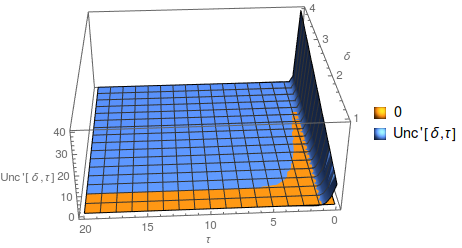}
 \caption{\label{deltar} {Plot of the time derivative of the uncertainty function w.r.t $\tau$, for fixed $\theta=1$, $\tau_0 =1$. } The first derivative test shows that the minimum in the uncertainty function appears for only a specific range of $\delta$ values, $1.367<\delta<3.890$, represented by the 2nd vanishing point of the first derivative of the uncertainty function w.r.t $\tau$.}
 \end{figure}\par 
In Figure \ref{delta2}, we show the results of numerical simulations for $\delta=2$, which shows a clear dip in the uncertainty function.
 \begin{figure}[htb!]
 \centering
 \includegraphics[scale=0.3]{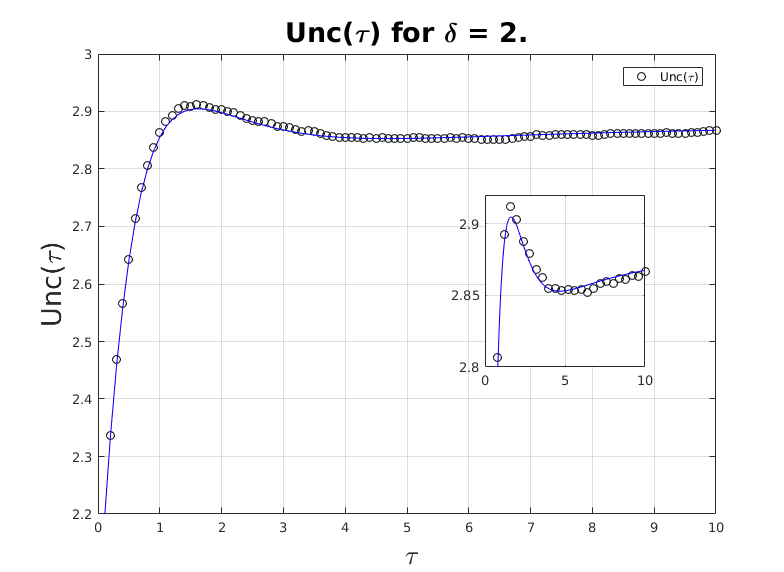}
 \caption{\label{delta2} Plot of the uncertainty function for $\delta= 2$. The symbols show results from the simulation of the Langevin equation with a step size of $ \Delta t=0.001$ and an average over $10^6$ realizations. We set $D=1$, $\theta =1$ and $\tau_0=2$. The lines correspond to exact solutions. }
 \end{figure}
In Figure \ref{tmparameter}, We further analyse the dependence of $\tau_m$ on the parameters $\theta$ and $\tau_0$. For fixed values of $\delta$ and $\tau_0$, $\tau_m$ is found to be independent of $\theta$. For $\delta = 2$ and $\tau_0 =1$, we find $\tau_m = 2.34$. However, $\tau_m$ is seen to increase linearly with $\tau_0$ for fixed values of $\delta$ and $\theta$. \par
 \begin{figure}[htb!]
  \centering
        \begin{subfigure}{0.5\textwidth}
        \centering
        \includegraphics[scale=0.5]{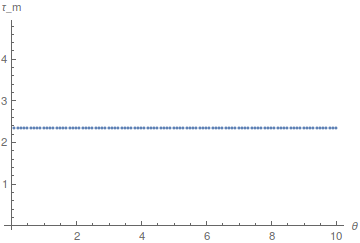}
        \caption{\label{delr} $\tau_m$ vs. $\theta$ for $\delta=2$ and $\tau_0 =1$}
    \end{subfigure}%
        \begin{subfigure}{0.5\textwidth}
        \centering
        \includegraphics[scale=0.5]{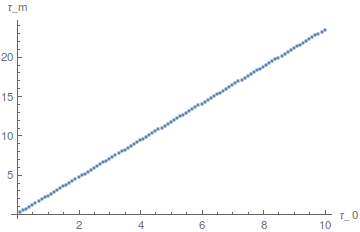}
        \caption{$\tau_m$ vs. $\tau_0$ for of $\delta=2$ and $\theta =1$}
    \end{subfigure}
    \caption{\label{tmparameter}The dependence of $\tau_m$ on the parameters $\theta$ and $\tau_0$, for a fixed value of $\delta$.}
\end{figure}
\section{Breathing parabola}
\label{bpunc}
Here we analyse the behaviour of a colloidal particle in a \textit{breathing parabola} potential. This is interesting to do since its a solvable yet non-trivial example, of a system
which never reaches a steady-state. The potential is given by the expression
\begin{equation}
V(x(t),\lambda(t))=\frac{\lambda(t)}{2}\;x(t)^2,
\end{equation}
where the stiffness of the trap, $\lambda$ is changing in a time-dependent manner according to a given protocol. The dynamics of the colloidal particle can be described using the Langevin equation,
\begin{equation}
\dot{x}(t)=-\partial_xV(x(t),\lambda(t
)) +\sqrt{\frac{2}{\beta}}\circ \eta(t).
\end{equation}
The dissipated work for this system for the equilibrium initial conditions,
\begin{align}
P_{\lambda_0}(x_0)&=\frac{1}{Z_0}e^{-\beta V(x_0,\lambda_0)},&Z_0&=\int_{-\infty}^\infty e^{-\beta V(x_0,\lambda_0)} dx_0.
\end{align}
 is given by,
\begin{align}
W_d[x(\cdot)]&=W[x(\cdot)]-\Delta F, & W[x(\cdot)]&=\int_0^\tau \frac{\partial V}{\partial \lambda }\dot{\lambda},& \Delta F &= \frac{1}{2}\log \frac{\lambda(\tau)}{\lambda(0)}.
\end{align}
Here, we will consider the specific forward protocol,
\begin{equation}
\label{bretpf}
\lambda^F(t)=\frac{1}{1+\tau-t},
\end{equation}
and the corresponding reverse protocol,
\begin{equation}
\label{bretpr}
\lambda^R(t)=\frac{1}{1+t}.
\end{equation} 
The methods discussed for the SSP are applicable to this simpler one-dimensional system as well \cite{skm}. For the forward process a closed form for the MGF of $W$ can be computed,
\begin{align}
\Phi^F(u)&=\scalebox{1} {$\sqrt{\frac{2(4 u+1) (\tau+1)^{\frac{1}{2} \left(\sqrt{4 u+1}+1\right)}}{\sqrt{4 u+1} (\tau+1)^{\sqrt{4 u+1}}+(\tau+1)^{\sqrt{4 u+1}}+u \left(\sqrt{4 u+1} (\tau+1)^{\sqrt{4 u+1}}+4 (\tau+1)^{\sqrt{4 u+1}}-\sqrt{4 u+1}+4\right)-\sqrt{4 u+1}+1}}$}
\end{align}
The MGF for the reverse protocol can also be computed using the same methods or by using the symmetry relation 
\begin{equation}
 \Phi^F(u)=e^{-\Delta F}\Phi^R(2- u).
\end{equation} 
  In Fig. \ref{mgfbp}, we plot the MGF of $W$ for both the forward and reverse protocols.
\begin{figure}[htb!]
 \centering
 \includegraphics[scale=0.5]{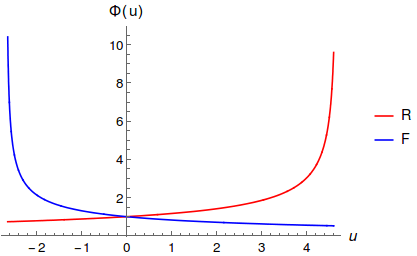}
 \caption{\label{mgfbp} Plot of the MGF of $W_d$ for the breathing parabola problem, for the forward (F) and reverse (R) protocols, for $\tau=1$. The MGFs satisfy the symmetry relation $\Phi^F(u)=e^{-\Delta F}\Phi^R(2- u)$.}
 \end{figure}
Using the exact form of the MGF, the moments of $W$ as well as the uncertainty function, can be computed at arbitrary times $\tau$. We find for the forward process,
\begin{equation}
\text{Unc}^F(\tau)=\frac{8 \left(\tau^2+3 \tau+2\right) \log (\tau+1)-\tau (15 \tau+16)}{(\tau+1) ((\tau+1) \log (\tau+1)-\tau)}.
\end{equation}
It can be shown that Unc$(\tau) \rightarrow 2$ as $\tau \rightarrow 0$ and Unc$(\tau) \rightarrow 8$ as $\tau \rightarrow \infty$. Similarly for the reverse process we find,
\begin{equation}
\text{Unc}^R(\tau)=-\frac{\tau \left(\tau^2+3 \tau+3\right) \left(\tau \left(\tau^2+3 \tau+3\right)+24 (\tau+1)^3 \log (\tau+1)\right)}{9 (\tau+1)^3 \left(\tau \left(\tau^2+3 \tau+3\right)+9 (\tau+1)^3 \log \left(\frac{1}{\tau+1}\right)+6 (\tau+1)^3 \log (\tau+1)\right)},
\end{equation}
which $\rightarrow 2$ as $\tau \rightarrow 0$ and $ \rightarrow 8/9$ as $\tau \rightarrow \infty$.
Unc($\tau$) for both the forward and reverse processes are plotted in Fig. 6 of the main text.
\end{widetext}

\end{document}